\documentclass[twocolumn,showpacs,preprintnumbers,amsmath,amssymb,floatfix]{revtex4}

\usepackage{graphicx}
\usepackage{dcolumn}
\usepackage{bm}

\begin{document}

\preprint{APS/123-QED}

\title{Neutral Evolution as Diffusion in phenotype space: reproduction with mutation but without selection}

\author{Daniel John Lawson}
\email{daniel.lawson@imperial.ac.uk}
\author{Henrik Jeldtoft Jensen}
\email{h.jensen@imperial.ac.uk}
\affiliation{
Department of Mathematics, Imperial College London\\
South Kensington campus, London SW7 2AZ
}

\date{\today}

\begin{abstract}
The process of `Evolutionary Diffusion', i.e. reproduction with local mutation but without selection in a biological population, resembles standard Diffusion in many ways. However, Evolutionary Diffusion allows the formation of localized peaks that undergo drift, even in the infinite population limit. We relate a microscopic evolution model to a stochastic model which we solve fully. This allows us to understand the large population limit, relates evolution to diffusion, and shows that independent local \emph{mutations} act as a \emph{diffusion} of interacting particles taking larger steps.
\end{abstract}



\maketitle

Reproduction involving random mutations seems at first to lead to a diffusion
of the population in type space,
however the diffusion involved is anomalous in various ways. 
A localized configuration that we call a `peak' forms in type space
\cite{FitnessLandscapes,Forster05-Quasispecies}, and
diffuses as a single entity.  The variations in the peak width increase
as the peak width itself with increasing population size, rendering
the infinite population limit meaningless.  In contrast, the distribution
of a large number of non-interacting particles undergoing local diffusion
forms a Normal Distribution with width increasing in time.
We will argue that a completely solvable 
stochastic differential equation model
captures the same dynamics as the microscopic evolution process, and
provides a meaningful description for the large population limit.
We show that although \emph{mutations} are independent, the effective 
\emph{diffusion} is not.

Much previous work on the clustering of individuals in type space focuses
on the genealogical lineage.  Ref. \cite{DerridaPeliti91} provides a 
comprehensive discussion and a complete solution from this viewpoint. 
We imagine a population of fixed size
$N$, and in each generation, some individuals can expect to have many
offspring and others will have none.  After some time the whole
population will have the same common ancestor, by the process of
Gamblers ruin\cite{BaileyStochastic}, and hence must have similar type.

Lineage analysis is a good tool to study high dimensional genotype spaces.
The theory of Critical Branching Processes\cite{Slade02-SuperBrownian}
finds that in high dimensions 
($d > d_c$, where the critical dimension $d_c=2$ \cite{Winter02-Branching})
describing genotype space, 
birth/death dynamics are described fully by the lineages.  
A lineage remains distinct until all individuals in it die.
However, in low 
dimensions ($d \leq d_c$) describing phenotypes, additional clustering
within a distribution occurs.
Although sometimes distinct, the clusters in phenotype
space can merge, and hence clusters are poorly defined 
entities.  Instead, a careful average over the distribution that we call 
a `peak' provides a more useful description.
Low dimensional clustering due to birth-death processes
was previously only understood in real space
\cite{ZhangEtAl90-Diffusion,Meyer96-clustering},  
with neutral phenotype clustering addressed indirectly
\cite{KesslerEvSmooth97,OhtaKimura73}.

The clustering described above is fluctuation driven.
Fluctuations must be considered in evolution unless the number 
of individuals per type is
high\cite{traulsen06-CoevFinitePop}, or there is strong selection
\cite{Zhang97-Quasispecies}.
Otherwise, \emph{there is always a region in
type space in which the population is small}, and therefore there is
an area of the equilibrium distribution that is affected by
noise.  It is (only) in the fluctuations that Evolutionary Diffusion
differs from normal Diffusion.

Understanding neutral evolution (i.e. reproduction with mutation but
without selection) is of great importance due to its
wide usage in numerous contexts, from Genealogical Trees
\cite{ServaLackSelfAveraging,ServaNeandertal,RannalaYangTrees96}, to
models of mutations in RNA \cite{KimuraMolecularNeutral,Bastolla02}. 
Neutral models
provide good matches with observed Species-Area Relations and
Species-Abundance Distributions\cite{NeutralTheory}.


\emph{Microscopic model:} We are interested in the 
distribution of types in
a population of individuals as they evolve. 
For comparison to Diffusion, we assume that the total
population $N(t) = N$ is constant, a restraint that can easily be relaxed.
In addition, we use the simplest type space, namely the 1-dimensional
set of integers.  However, the qualitative behavior discussed will
remain the same in all large connected type spaces.  The timestep for
the \emph{microscopic} processes we consider are:


\emph{The Diffusion Process}:\vspace{-5pt}
\begin{enumerate}
\item Select an individual $i$ (at position $x$), each with
    probability $1/N$. \vspace{-5pt}
\item Move to $y = x \pm 1$ each with probability $p_{m}/2$, or
  leave at $y=x$ with probability $1-p_{m}$.\vspace{-5pt}
\end{enumerate}

\emph{The Evolutionary Diffusion process} (which is the Moran process \cite{Moran62} 
for a type distribution):\vspace{-5pt}
\begin{enumerate}
\item Select an individual $i$ (at position $x$), each with
  probability $1/N$  and mark for killing.\vspace{-5pt}
\item Select an individual $j$ (at position $x_j$) for reproduction,
  each with probability $1/N$. \vspace{-5pt}
\item  Remove individual $i$, and create an offspring
  of individual $j$ at $y = x_j$ with
  probability $(1-p_{m})$, or mutate to $y = x_j \pm 1$ each with
  probability $p_{m}/2$.  Hence the effective diffusive step is
  $y-x$.
\end{enumerate}
We will refer to properties of the Diffusion process with the
subscript $D$, and the Evolutionary Diffusion process with the
subscript $E$, e.g. ${\langle x \rangle}_E(t)$ for the mean position
of the individuals in the evolution process after $t$ timesteps.  
Time is best measured in \emph{generational time} $T = t/N$.
Care is needed when averaging: we will use 
the ensemble average (over many realizations) of a quantity $V$
$\overline{V}(t)$, population average 
$\langle V \rangle(t) = \sum_i^N V_i(t)/N$ and time
average up to time $\tau$: $\langle V \rangle = \sum_{t=t_0}^\tau V(t)/(\tau-t_0)$.  
Quantities calculated from probabilities
are by definition ensemble averages, and so the notation refers to which
average is taken first.  See \cite{DerridaPeliti91} for further details.

\begin{figure}[ht]
  \centering
  \includegraphics[width=70mm]{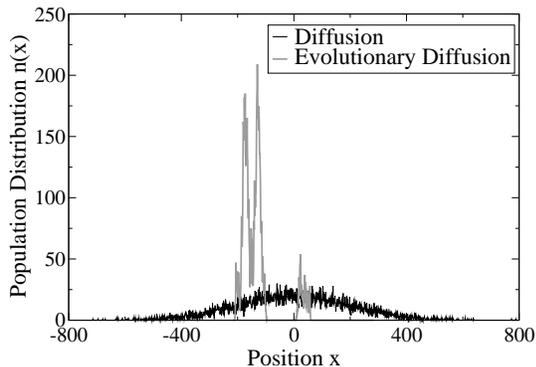}
\caption{A snapshot of the distribution after $80000$ generations, using
$N=10000$ and $p_{m}=0.5$, comparing Evolutionary Diffusion (grey
line) and Diffusion (black line).  Diffusion
follows a (noisy) Normal distribution whereas Evolutionary Diffusion
is localized as  $2$ clusters, which we combine as a 'peak' of width $w$ 
and position $\mu$ undergoing drift. \label{fig:sampledistrib}}\vspace{-10pt}
\end{figure}

The number of individuals on site $x$ is $n(x,t)$,  and the initial
conditions are $n(x=0,t=0)=N$, $n(x,t)=0$ for $x\ne 0$. The ensemble average
of the population distribution $\overline{n}(x,t)$ is 
obtained directly from the Master equation, and is identical for both
Diffusion and Evolutionary Diffusion:
\begin{equation}
\frac{\Delta \overline{n} (x,t)}{\Delta t} = \frac{p_{m}}{2N} \bigtriangledown^2
\overline{n}(x,t).
\end{equation}
Hence the (one-point) ensemble average of the two processes is
the same, but numerical simulations
(Fig. \ref{fig:sampledistrib}) reveals very different behavior.  From
the figure, we see that Diffusion has followed the ensemble average: a
Normal distribution centered on $0$, increasing in width with
time\cite{BaschnagelStoch99}.  
Although we shall see that the Evolutionary Diffusion process 
self-averages over time, the thermodynamic limit is subtle.
In order to understand why, we now split the peak
up into its mean position and standard deviation to create a ``Theory of
evolutionary peaks''.

\begin{figure}
[htb]
\includegraphics[width=65mm]{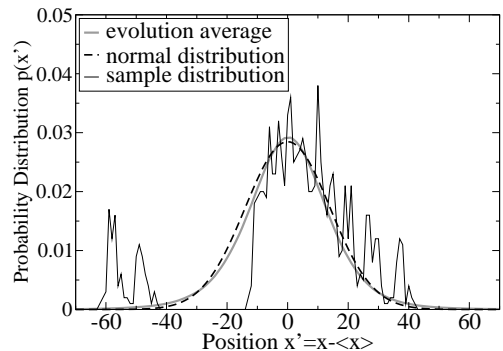}
\caption{Time-averaged Evolutionary Diffusion distribution (solid
line), Normal distribution (dashed line) with standard deviation
calculated from theory in Eq. (\ref{eqn:meanw}).  The two 
agree up to the second moment.
Also shown is a snapshot of
the distribution (thin line). ($N=1000$, $p_{m}=0.5$)
\label{fig:timeavedistrib}}
\vspace{-10pt}
\end{figure}

\emph{Theory of evolutionary peaks:} We define here conceptually 
simple and solvable processes 
of Evolutionary Diffusion and Diffusion which
we argue captures the essential features of the microscopic models.  
The distribution is described as a `peak':
a Normal distribution
with mean $\mu(t)$ and standard deviation (i.e. width) $w(t)$, 
which vary as a product of the dynamics.  The probability
distribution is continuous, but a discrete `individual' of size $1/N$
is moved per timestep.  Although a given realization of a peak never
resembles a Normal distribution, this is a good model of the
evolutionary process because a Normal distribution is a good approximation
for the time average of the peaks in the variable 
$x' = x - \mu(t)$ (we now drop the dash notation); see Fig. \ref{fig:timeavedistrib}.
We hence `integrate out' the \emph{inessential} degrees 
of freedom: the particular distribution of individuals within the peak.

In the evolutionary process, in each timestep a death will occur at
any point $x$ in the distribution $p(x)$:
\begin{equation}
p_E(x;\mu=0,w) = \frac{e^{-x^2/2w^2}}{\sqrt{2 \pi} w}. \label{eq:pkE}
\end{equation}
The parent position $x_j$ will be drawn independently from the same
distribution, and the offspring will be mutated with probability
$p_{m}$ to $ y \pm 1$.  Hence the distribution for births $p(y)$ is:
\begin{align}
p_E(y;\mu=0,w) &=(1-p_{m})\frac{e^{-y^2/2w^2}}{\sqrt{2 \pi} w}  \label{eq:poE}\\
&+\frac{p_{m}}{2}\Big[\frac{e^{-(y-1)^2/2w^2}}{\sqrt{2 \pi} w} +
\frac{e^{-(y+1)^2/2w^2}}{\sqrt{2 \pi} w}\Big].  \nonumber
\end{align}
The probability distribution for the Diffusion process, moving a
particle at $x$ to $x \pm 1$ with probability $p_{m}$, is written as:
\begin{align}
p_D(x;\mu=0,w) =& \frac{e^{-x^2/2w^2}}{\sqrt{2 \pi} w}, \\ 
p_D(y;\mu=0,w) =&
(1-p_{m})\delta(y-x) \\
&+\frac{p_{m}}{2} [\delta(y-x+1) + \delta(y-x-1)].\nonumber
\end{align}
The expectation value of a variable $V(x,y)$ is simply the integral of
$V$ over the probability distribution:
\begin{equation}
\langle V(x,y) \rangle = \int_{-\infty}^{\infty}
\int_{-\infty}^{\infty} V(x,y) p(x) p(y) dy dx. \label{eq:expV}
\end{equation}
Eq. (\ref{eq:expV}) is simple to calculate because all
of our probabilities are independently Normal distributed, or
interact trivially via delta functions.

We now perform calculations for the expectation values of
$w(t+1)$ given $w(t)$, (working with the variance $w^2$
for simplicity).  We consider the death of individual $q$ at $x_q$,
which is replaced by a birth occurring at $y_q$.
\begin{align}
w^2(t) &= \frac{1}{N} \sum_{i=1}^N x_i^2 - \left(\sum_{i=1}^N
\frac{x_i}{N}\right)^2, \\ 
w^2(t+1) &= \frac{1}{N} \sum_{i=1}^N x_i^2
+ \frac{y_q^2 - x_q^2}{N} - \left( \sum_{i=1}^N \frac{x_i}{N} + \frac{y_q-x_q}{N}\right)^2, 
\\ 
F &= \Delta w^2(t) = w^2(t+1) - w^2(t) \nonumber \\
&=\frac{y_q^2-x_q^2}{N} - \frac{y_q^2 + x_q^2 - 2x_q y_q}{N^2}.\label{eqn:Fdef}
\end{align}
Here we have defined $F= \Delta w^2(t)$ for later use, and used
$\sum_{i=1}^N x_i = 0$.  These quantities are population averages; we 
now ensemble average over the possible births and deaths by simple 
integration over Eq. (\ref{eq:expV}). We find
that for the Diffusion process, the expected change in the variance is 
always positive and independent of $w$:
\begin{equation}
\frac{\Delta \langle w_D^2(x,y) \rangle}{\Delta t} 
= \frac{p_{m}}{N} (1-\frac{1}{N}). \label{eqn:diffdeltaw}
\end{equation}

For the evolution process, the expected change in the variance is:
\begin{equation}
\frac{\Delta \langle w_E^2(x,y) \rangle}{\Delta t} = \frac{1}{N}
(p^*-\frac{2 w_E^2}{N}), \label{eqn:dsigsqdt}
\end{equation}
where for brevity we have defined $p^* = p_m-1/N$ (assumed positive).
This time, the rate of change of the variance depends on itself, and
there is an equilibrium for which $E( \Delta w_E^2(x,y)) =0$, at
$w_E^{equil} = \sqrt{{N p^*}/{2}}.$
The product $Np^*$ is the average number of mutants per generation, 
minus one.  By taking the limit $\Delta t \to 0$ in
Eq. (\ref{eqn:dsigsqdt}), and solving by separation of variables, we
obtain the variance 
$\langle w_E^2 \rangle (T) = {\frac{N p^*}{2}} (1-e^{-2T/N})$.

We now look at how the peak width $w$ varies in time, by
considering the \emph{fluctuations} in $F = \Delta_t w^2$, the
change of peak size.  We are interested in fluctuations around the
equilibrium standard deviation $w^{equil}$.  
$w^{equil}$ is not the mean observed value of $w$ - we will
be able to correct it by considering higher moments.  We will now
assume a large population $N \gg 1$, and consider the reduced variable
$s = \frac{w}{\sqrt{N}}$ to identify leading order terms.
\begin{align}
\overline{F^2} - \overline{F}^2 
&= 4 s^4 + {4 p_{m} s^2}/{N} + \cdots \nonumber\\ 
& \approx \frac{4 w^4}{N^2}.
\label{fig:deltaF2}
\end{align}
To represent the particular history of the evolution process
we must write Eq. 
(\ref{eqn:dsigsqdt}) with an additional noise term 
$\sqrt{\overline{F^2} - \overline{F}^2} \eta(t) \approx (2w^2/N) \eta(t)$,
where $\eta(t)$ has mean zero and standard deviation $1$ (keeping up 
to second order moments in the noise - higher moments are $O(1/N)$ smaller).
In generational time $T=t/N$, as $\Delta T \to 0$ we obtain:
\begin{equation}
d w_E^2(T) \approx \left[p^* - \frac{2 w^2}{N}
\right] dT + \frac{2 w^2}{\sqrt{N}} dW . \label{eqn:itoeq}
\end{equation}
Where $W(t)$ is a Wiener process\cite{BaschnagelStoch99}.
We solve by finding the Fokker-Planck equation \cite{KannanStoch}:
\begin{align}
&&\frac{\partial p(w^2,T)}{\partial T} + \frac{\partial ([p^*-2w^2/N]p(w^2,T))}
{\partial (w^2)} \nonumber\\
&&- \frac{1}{2} \frac{\partial^2(4w^4 p(w^2,T) /N)}{\partial (w^2)^2} =0.
\end{align}
Seeking the steady state solution $\frac{\partial p(w^2,T)}{\partial T}=0$,
integration twice shows that (for this to be a probability distribution) the 
unique solution is:
\begin{align}
&p(w^2)d(w^2) = \left(\frac{N p^*}{2}\right)^2 \frac{1}{(w^2)^3}e^{-\frac{N p^*}{2w^2}} d(w^2),\\
&\implies \, p(w)dw = \frac{(N p^*)^2}{2} \frac{1}{w^5} e^{-\frac{N p^*}{2w^2}}dw. \label{eqn:probw}
\end{align}
The tail of $p(w)$ is a power law, corresponding to the existence of 
multiple (arbitrarily distant) clusters within the peak.
From this we can calculate the arithmetic mean of the peak width, corrected
for noise:
\begin{equation}
\langle w \rangle = \int_0^\infty w p(w) dw = 
\sqrt{\frac{N p^*}{2}} \frac{\sqrt{\pi}}{2}, \label{eqn:meanw}
\end{equation}
This contrasts with Diffusion, as $\langle w_D \rangle$ has no stationary 
distribution and follows Eq. (\ref{eqn:diffdeltaw}).  The standard deviation of the peak 
width is:
\begin{equation}
\sigma_w = \sqrt{\langle w^2 \rangle - \langle w \rangle ^2} = \sqrt{{N p^*(1-\pi/4)}/{2}}.
\end{equation}
Therefore the standard deviation in the peak width increases at the same rate 
(with N) as the peak width itself.  The 4th and higher moments of the distribution 
of peak widths diverge due to the power law tail of $p(w)$.  The model 
approximations are confirmed 
by numerics. Both Eq. (\ref{eqn:itoeq}) and $w(t)$ for the `Evolutionary
Diffusion Process' defined initially have indistinguishable signals and 
Power Spectra (not shown), and conform to Eq. (\ref{eqn:meanw}) to within $2\%$: 
for $N=10000$ and $p_{m}=1$, with $200$ runs of $10^5$ generations,
counting $w(t)$ after time $5 \times 10^4$, we find $\langle w \rangle= 64.34 \pm 2.14$ 
for the Evolutionary Diffusion
Process, $\langle w \rangle= 63.17 \pm 1.20$ for Eq. (\ref{eqn:itoeq}), comparing with a
theoretical prediction of $\langle w \rangle =62.66$.  Eq. (\ref{eqn:itoeq}) is fast to
simulate for long times and, as indicated, behaves
very similarly to the microscopic process.

We now examine the behavior of the expected
root-mean-square (RMS) displacement of the peak center as a function
of time; direct integration of $\langle(\Delta x)^2\rangle =
\langle(x_q-y_q)^2/N^2 \rangle$, using the steady state value $\langle w^2_E \rangle = N p^*/2$ in Eq. (\ref{eqn:xevsoln}), yields the following step size
for evolution:
\begin{equation}
\Delta \langle {x\rangle}_E^{RMS}(t) \approx \sqrt{{p^*}/{N}}.
\end{equation}
From random walk theory\cite{BaschnagelStoch99}, the mean (RMS)
position of a random walker taking steps of size $S$ after $t$ timesteps is
$\langle x \rangle^{RMS} = S \sqrt{t}$.  Hence:
\begin{align}
&{\langle x \rangle}^{RMS}_D(T) = {\sqrt{p_{m} t}}/{N} = \sqrt{{p_{m} T}/{N}},\\
&{\langle x \rangle}^{RMS}_E(T) = \sqrt{{p^* t}/{N}} = \sqrt{p^* T}.
\end{align}
Hence, in the limit of infinite $N$ the Diffusion process remains stationary,
but in generational time the mean position of the Evolutionary Diffusion
process does a random walk of step size \emph{independent of the
total number of individuals}.

For completeness we could write an
equation for $\mu(T)={\langle x \rangle}$ for evolution as:
$d \mu_E(T) = N^{-1/2}\sqrt{p_m + 2 w_E^2(t)} dW$.
This equation together with Eq. \ref{eqn:itoeq} describe the system fully
and are completely solved once the peak width reaches
equilibrium probability distribution.

We have described the microscopic behavior
of the evolution of reproducing individuals in a type space,
and approximated it to two coupled solvable stochastic processes for the distribution.
We find two main differences between Evolutionary Diffusion and
normal Diffusion. 1) The short range mutation process effectively
becomes a longer ranged (by O($\sqrt{N}$)) diffusive step. By the
Central Limit Theorem, the standard deviation of the mean position $\mu$
taking $N$ steps per generation of size $A$ increases as $A \sqrt{N}$.
In diffusion, the steps are of size $A=1/N$, but in evolution the steps are
of size $A=1/\sqrt{N}$ so the convergence is not
fast enough to set the location of the peak center in the infinite
population limit.
2) The effective diffusion is not independent and peaks can
form with fluctuating width $w$ around $\langle w \rangle$,
following the distribution in Eq. (\ref{eqn:probw}) which has a power law tail.  
This provides a null hypothesis to determine if two asexual individuals belonging to different clusters of a phenotype in fact are subject to the \emph{same} selection pressure, i.e. members of a single \emph{neutrally} evolving population or 'peak', or whether differential selection is responsible for the population breaking up into separate clusters. In the neutral case all but one cluster will go extinct. However, if differential selection acts then several clusters of phenotype may survive in separate `niches'.

In terms of replicator dynamics, our results transparently 
explain how a `species' in type space (the peak 
described above) is able to maintain its coherence as it performs a random
walk due to mutation prone reproduction.  We found that the distribution of a phenotype
in neutral evolution is `non-trivial' regardless of population size.
In terms of diffusion, we
describe an interesting type of particle interaction that allows for clustering.

DL is funded by an EPSRC studentship, and is grateful to
Nelson Bernardino for useful discussions.

\end{document}